\documentclass[preprint,showpacs,preprintnumbers,amsmath,amssymb,aps]{revtex4}
\usepackage{graphics}% Include figure files
\usepackage{graphicx}% Include figure files
\usepackage{dcolumn}% Align table columns on decimal point
\usepackage{bm}% bold math

\draft % marks overfull lines with a black rule on the right

\begin{document}

%\preprint{}

\title{Andreev-reflection spectroscopy of ferromagnets: 
  the impact of Fermi surface mismatch} %Title of paper



\author{Elina Tuuli{\email{estuul@utu.fi}}$^{1,2,3}$ and Kurt Gloos$^{1,3}$}
%\thanks{}
\affiliation{$^1$Wihuri Physical Laboratory, Department of Physics and Astronomy, 
  University of Turku, FIN-20014 Turku, Finland}
\affiliation{$^2$Graduate School for Materials Research,
  (GSMR), FIN-20500 Turku, Finland}
\affiliation{$^3$Turku University Centre for Materials and Surfaces 
  (MatSurf), FIN-20014 Turku, Finland}

\email{estuul@utu.fi}


\date{\today}

\begin{abstract}
We have investigated point contacts between a superconductor (Nb, AuIn$_2$) 
and a normal metal (ferromagnetic Co, non-magnetic Cu). 
The observed Andreev-reflection spectra were analyzed using the modified BTK 
theory including spin-polarization effects. 
This resulted in a polarization of Co that agrees with observations by others, 
but lifetime effects describe the spectra equally well. 
On the other hand, the spectra with non-magnetic Cu can be well described 
using the spin-polarization model. 
The ambiguity between polarization and lifetime interpretation poses a dilemma 
which can be resolved by considering the normal reflection at those interfaces 
due to Fermi surface mismatch. 
Our data suggest that Andreev reflection at Nb - Co contacts does  deliver the 
true magnetic polarization of Co only when lifetime effects and the above 
mentioned intrinsic normal reflection are included.

\end{abstract}


% 74.45.+c 	Proximity effects; Andreev reflection; SN and SNS junctions
% 72.25.Mk 	Spin transport through interfaces 
% 73.40.-c 	Electronic transport in interface structures
% 85.30.Hi 	Surface barrier, boundary, and point contact devices 
% 75.50.Cc 	Other ferromagnetic metals and alloys 


\maketitle %\maketitle must follow title, authors, abstract and \pacs

%#########################################################################
%#########################################################################
\section{Introduction}

Andreev-reflection spectroscopy at point contacts has been suggested as a 
versatile tool to determine the magnetic (spin current) polarization $P$ of 
ferromagnets \cite{deJong1995,Soulen1998}. 
Today it is widely believed 
\cite{Upadhyay1998,Ji2001a,Kant2002,Auth2003,Bugoslavsky2005,Mukhopadhyay2007,%
Chalsani2007,Stokmaier2008,Baltz2009}
that the polarization can be reliably extracted from the measured 
point-contact spectra by applying a modified version of the BTK theory of 
Andreev reflection \cite{Blonder1982}, like Strijkers' \cite{Strijkers2001} or 
Mazin's model \cite{Mazin2001}. 
However, it has also been noted that the interpretation of these point-contact 
spectra presents extra difficulties because of spurious  superposed anomalies
and the poor convergence of the fitting procedure
\cite{Bugoslavsky2005,Chalsani2007,Baltz2009}.

Andreev reflection across a ballistic contact between a normal metal and 
a superconductor requires the transfer of an electron pair with opposite 
momentum and spin from the normal conductor to form a Cooper pair in the 
superconductor.
In an equivalent description an electron is transferred to the superconductor 
and the corresponding hole is retro-reflected.
This reduces the contact resistance by a factor of two for energies within the 
superconducting gap $2\Delta$. 
Normal reflection at the interface has a pronounced effect on the shape of the 
spectra because it enters the pair transfer twice by affecting the incident 
electron and also the retro-reflected hole, yielding the typical  double-minimum
structure of Andreev reflection. 
To keep the number of adjustable parameters as small as possible, Blonder, 
Tinkham, and Klapwijk  \cite{Blonder1982} described normal reflection by a 
$\delta$-function barrier of strength $Z$. 
The BTK theory is well accepted to analyze the Andreev-reflection spectra of 
ballistic contacts between BCS-type superconductors and non-magnetic normal 
metals.

Since interfaces are usually not perfect, they cause additional scattering 
that can break up the Cooper pairs and, thus, reduce the superconducting  
order parameter.
Dynes' model describes this situation by a finite lifetime 
$\tau = \hbar/\Gamma$ of the Cooper pairs, which strongly  reduces the 
magnitude of the  Andreev-reflection anomaly \cite{Plecenik1994}.

A magnetically polarized metal has an unequal number of spin up and spin 
down electrons. 
Conduction electrons  that can not find their corresponding pair with opposite 
spin do not participate in Andreev reflection, opening the way to directly 
measure the polarization \cite{deJong1995,Soulen1998}.
The polarization reduces the magnitude of the Andreev-reflection anomaly
like the lifetime effects, and it leads to a zero-bias maximum 
of the differential resistance similar to that of normal reflection. 
With few exceptions \cite{Mukhopadhyay2007,Chalsani2007},
the analysis of superconductor - ferromagnet spectra 
usually excludes lifetime effects
\cite{Upadhyay1998,Ji2001a,Kant2002,Auth2003,Bugoslavsky2005,%
Stokmaier2008,Baltz2009}
so that $P$ and $Z$ are the only main adjustable parameters. 
Also the superconducting energy gap has to be treated as a variable, although
its approximate value at the contact is known from the bulk superconducting 
properties.
Often a so-called 'broadening parameter' is included to improve the fit quality  
by simulating an enhanced smearing  of the Fermi edge 
\cite{Kant2002,Bugoslavsky2005,Baltz2009}, 
but increasing the  parameter number  means the solution 
becomes more easily degenerate.

We show here that the Andreev-reflection spectra of both
ferromagnets and non-magnets can be fitted equally well by assuming a magnetic 
polarization of the normal metal without lifetime effects and vice versa. 
This problem can be solved by taking into account the lower bound of $Z$ 
due to Fermi surface mismatch.

%#########################################################################
%#########################################################################
\section{Experimental}

Our experiments are based on  shear contacts   between superconducting Nb 
($T_c = 9.2\,$K) and normal conducting Co and Cu wires ($\sim 0.25\,$mm 
diameter) at $4.2\,$K in liquid helium.
Co is a band-ferromagnet with $T_{Curie} = 1388\,$K and  Cu a non-magnetic 
normal metal \cite{Ashcroft1976}.
We have also re-analyzed older spear-anvil type experiments at 
$0.05\,$K  with the BCS-type superconductor AuIn$_2$ ($T_c = 0.21\,$K) 
in contact with a Cu wire \cite{Gloos1996}. 
The differential resistance $dV/dI$ was measured as function of bias voltage
$V$ with low-frequency current modulation in four-wire mode. 

%=========================================================================
\begin{figure}
  \includegraphics{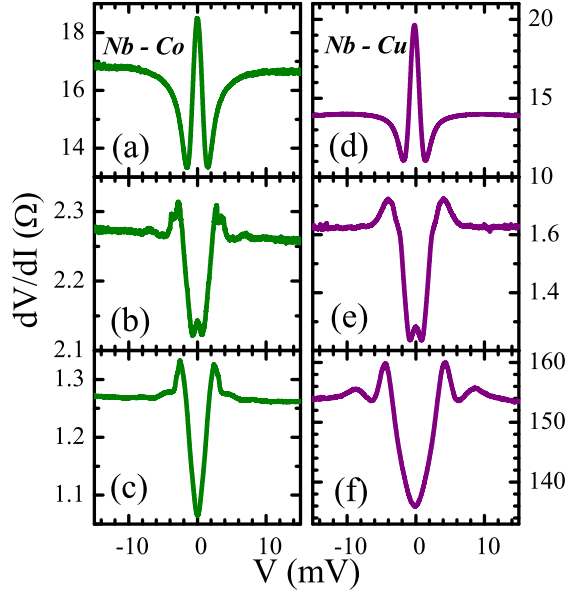}
  \caption{(Color online) Typical $dV/dI(V)$ spectra of (a - c) Nb - Co 
  and (d - f) Nb - Cu contacts at 4.2\,K.}
  \label{spectra} 
  \end{figure}
%=========================================================================

Figure \ref{spectra} shows  typical spectra of Nb - Co as well as Nb - Cu 
contacts. 
We have observed various types that can be classified as follows: 
$i)$ Andreev-reflection double minimum (a, d), 
$ii)$ Andreev reflection with side peaks (b, e), 
$iii)$ single zero-bias minimum with or without side peaks (c, f), and 
$iv)$ zero-bias maximum without signs of superconductivity (not shown). 
For our analysis we have used only contacts of type $i)$ and $ii)$ which show 
the 'hallmark' of Andreev reflection. 
The origin of the side peaks will be discussed elsewhere. 
Contacts of type $iv)$ were studied earlier \cite{Gloos2009}.

%#########################################################################
%#########################################################################
\section{Discussion}

\begin{figure}
  \includegraphics{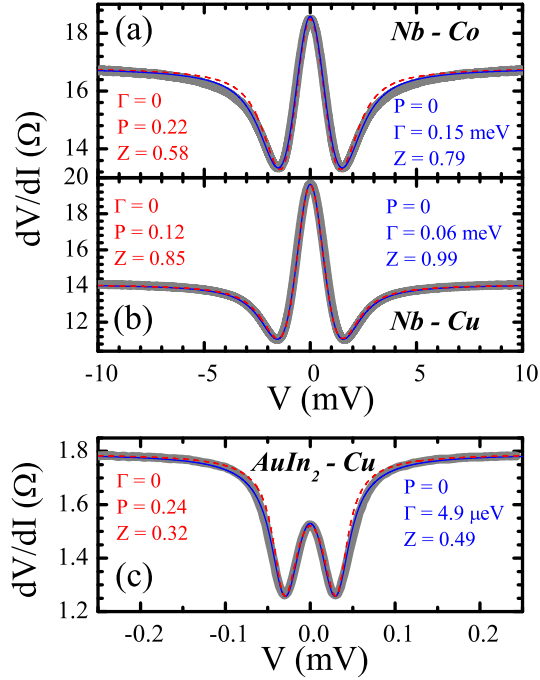}
  \caption{(Color online) 
  Typical differential resistance versus bias voltage (thick solid lines) 
  together with fits derived by assuming $\Gamma=0$ (thin dashed lines) and 
  $P=0$ (thin solid lines) using the indicated fitting parameters. 
  For all contacts the curves are almost   indistinguishable.
  Deviations found only near the shoulder where $dV/dI$ starts to drop from 
  its normal-state value can be removed by  introducing a 'broadening 
  parameter'.
  (a) Nb - Co at $T = 4.2\,$K and $2\Delta = 2.6\,$meV,
  (b) Nb - Cu at $T = 4.2\,$K and $2\Delta = 2.5\,$meV, and
  (c) AuIn$_2$ - Cu at $T = 0.05\,$K and $2\Delta = 65\,\mu$eV.}
  \label{Fits} 
  \end{figure}

\begin{figure}
  \includegraphics{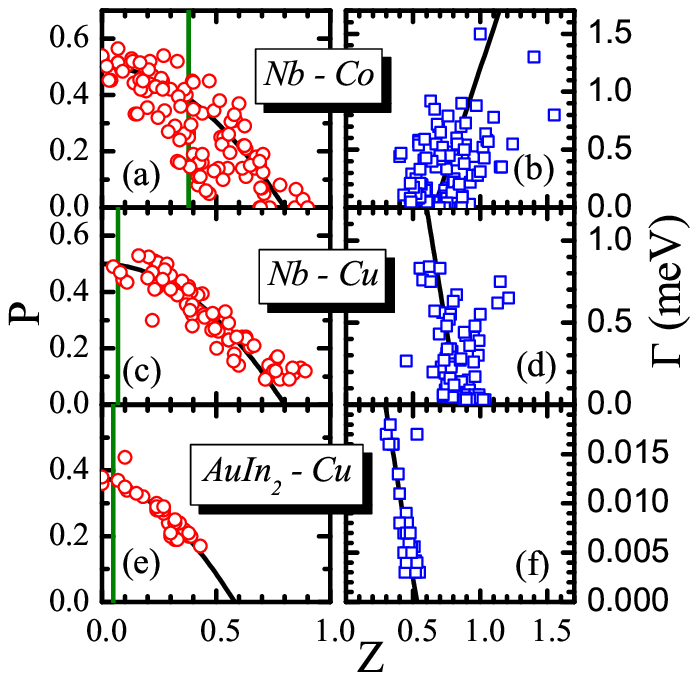}
  \caption{(Color online) 
  Polarization $P$ at $\Gamma = 0$ and life-time broadening $\Gamma$ 
  at $P = 0$   versus $Z$ of Nb - Co,   Nb - Cu, and AuIn$_2$ - Cu contacts. 
  The vertical solid lines represent the expected minimum $Z_0$ due to 
  Fermi momentum mismatch in free-electron approximation. 
  Solid lines through the data points serve as guide to the eye.}
  \label{PolGamma} 
  \end{figure}

We fitted the spectra in the conventional way 
\cite{Upadhyay1998,Ji2001a,Kant2002,Auth2003,Bugoslavsky2005,Mukhopadhyay2007,%
Chalsani2007,Stokmaier2008,Baltz2009} using Strijkers' model and 
assuming $\Gamma = 0$ (Figure \ref{Fits}). 
Mazin's model would only slightly change the $P(Z)$ data 
\cite{Ji2001b}. 
The resultant polarization of Nb - Co contacts in Figure \ref{PolGamma} (a)
agrees well with that found by others \cite{Strijkers2001,Kant2002,Baltz2009}.
However, analysing the Nb - Cu and the AuIn$_2$ - Cu spectra in the same way,
assuming $\Gamma = 0$ and allowing $P$ to vary, yielded almost the 
same $P(Z)$ as for the Nb - Co contacts (Figure \ref{PolGamma}):
Without advance knowledge that Cu has zero spin polarization $P=0$,
we would be led to believe that it is actually polarized like ferromagnetic Co. 
Such a possibility was mentioned -- but discarded -- by Chalsani 
{\it et al.} \cite{Chalsani2007} for Pb - Cu contacts.
Nevertheless, this speculation could be supported by recent experiments on the
size-dependence of the so-called zero-bias anomaly which has been attributed 
to the spontaneous electron spin  polarization at the point contact 
\cite{Gloos2009}.

%=========================================================================
%=========================================================================

It appears trivial to assume $P = 0$ for Cu and to use the lifetime parameter 
$\Gamma$ that fits the observed spectra equally well. 
But the lifetime-only model also works well for ferromagnetic Co, as
demonstrated in Figure \ref{Fits} where the theoretical curves for the two 
fitting procedures can be barely separated. 

%=========================================================================
%=========================================================================

\begin{figure}
  \includegraphics{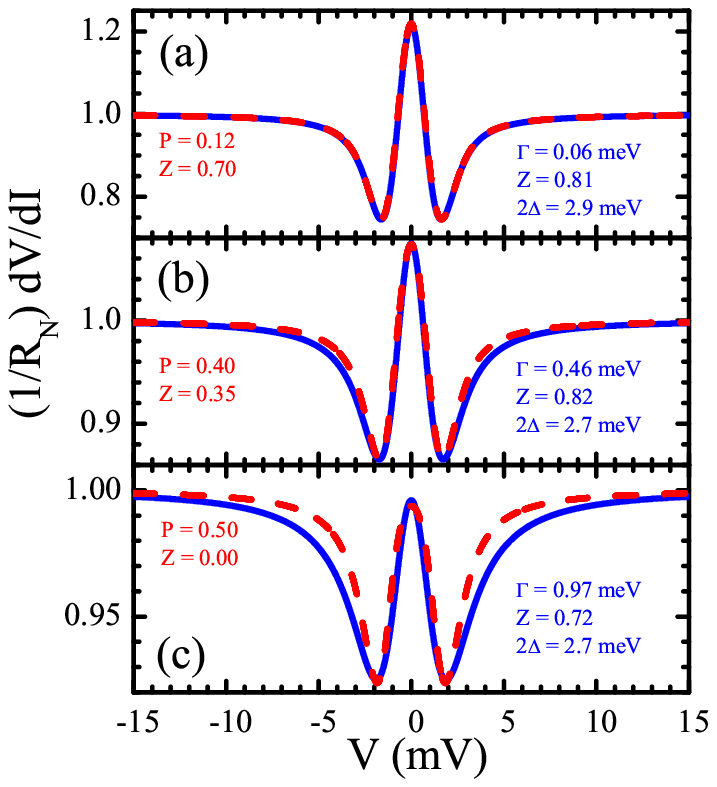}
  \caption{(Color online) 
  Comparison between the polarization-only (red dashed lines) and lifetime-only 
  (blue solid lines) models for contacts with (a) small, (b) medium, and (c) 
  large polarization. The differential resistance $dV/dI$ is normalized to the
  normal contact resistance $R_N$.
  First the polarization-only spectra were calculated assuming the indicated $P$
  and $Z$ at $2\Delta = 3.0\,$meV for niobium and $T=4.2\,$K. Then the 
  lifetime-only spectra were fitted, resulting in the indicated $\Gamma$ and $Z$. 
  For this fitting the energy gap had to be slightly adjusted.}
  \label{Models} 
  \end{figure}

In order to study the similarities and differences 
between the two models in more detail, we 
have calculated spectra at small, medium, and large values of $Z$ together 
with their typical polarization as found in the experiments 
summarized in Figure \ref{PolGamma}.
These theoretical curves were then fitted with the lifetime-only model. 
Figure \ref{Models} demonstrates the perfect agreement between the two models
at large $Z$ and small $P$. 
This confirms earlier findings by Chalsani {\em et al.} in the case of Pb - Cu
and Pb - Co contacts \cite{Chalsani2007}.
Deviations become obvious only at small $Z$ and large $P$. 
Note also that the strong $Z$-dependence of $P$ turns into a $\Gamma$
at nearly constant $Z$, in agreement with the experimental data 
in Figure \ref{PolGamma}.
Consequently, distinguishing lifetime effects from the magnetic polarization
requires additional information.

%=========================================================================
%=========================================================================

This knowledge could be obtained from normal reflection:
Figure \ref{PolGamma} shows that the $P(Z)$ data are almost evenly distributed
on the $Z$ axis from $Z \approx 0$ to the maximum value of $Z \approx 0.8$ for Nb-Co and
Nb-Cu contacts. 
In contrast, the $\Gamma(Z)$ data are centered at around $Z \approx 0.8$,
indicating a preferred value for normal reflection.
This different behaviour must have a reason.

$Z$ consists of two parts, $Z_{barrier}$ describes reflection at a 
possible interface tunneling barrier (and any other mechanism that might be 
subsumed under this term), and  $Z_0$ due a mismatch of the Fermi surfaces 
or band structures of the two electrodes. 
In free-electron approximation Fermi surface mismatch reduces to a mismatch 
$r = v_{F1} / v_{F2}$ of Fermi velocities $v_{F1,2}$ on both sides of the 
contact and results in \cite{Blonder1983}
\begin{equation}
  Z^2  = Z_{barrier}^2 + Z_0^2 = Z_{barrier}^2 + \frac{(1 - r)^2}{4r}.
  \label{Z-parameter}
  \end{equation}
Thus $Z_0$ defines a lower bound of $Z$ when a tunneling barrier is absent.
That means, without tunneling barrier the $Z$ parameter of the contacts for
a given metal combination should be constant while
a tunneling barrier would add a tail to the $Z$ distribution at large values.
The experimental data in Figure \ref{PolGamma} indicate that our contacts
either have a negligibly small $Z_0$ plus an irreproducible tunneling
barrier (polarization-only model) or a large $Z_0$ with a negligibly small 
tunneling barrier (lifetime-only model).

Note that Eq. \ref{Z-parameter} requires equal effective electron masses.
For example, Fermi velocity mismatch is negligible at interfaces between
a heavy-fermion compound and a simple metal because their huge velocity 
mismatch of up to $r  \approx 1000$ is compensated by the large
mismatch of the effective electron masses \cite{Deutscher1994}. 
Therefore it is more appropriate to speak of a momentum mismatch 
instead and replace the variable $r$ by the ratio of 
Fermi wave numbers $k_{F1, 2}$.

While we do not know whether our point contacts possess a tunneling barrier, 
it should be possible to predict $Z_0$ from the known band structure of metals. 
This turns out to be quite difficult because there are different theoretical
and experimental estimates for the Fermi surface properties.
%velocities $v_F$, effective electron masses $m$, and Fermi wave numbers $k_F$.
In free-electron approximation $k_F = 13.6\,$nm$^{-1}$ for Cu and 
$k_F = 11.8\,$nm$^{-1}$ for Nb \cite{Ashcroft1976}. 
AuIn$_2$ has nearly the same conduction electron density as Cu and, thus, 
a very similar $k_F$ \cite{Rayne1963}.
Co has spin-split energy bands, and therefore different wave numbers for the 
two spin directions. Its average Fermi velocity $v_F \approx 280\,$km/s is known from 
critical-current oscillations in Josephson $\pi$-junctions \cite{Robinson2006}. 
Its effective electron mass $m$ is about twice the free electron mass
\cite{McMullan1992}, yielding $k_F = mv_F/\hbar \approx 5.6\,$nm$^{-1}$.
The minimum $Z$ parameters $Z_0 \approx 0.05$ for AuIn$_2$ - Cu
\cite{Gloos1996}, $Z_0 \approx 0.07$ for Nb - Cu, and $Z_0 \approx 0.38$ for Nb - Co
are consistent with the polarization-only and with the lifetime-only
model  for Nb - Cu and AuIn$_2$-Cu, but they clearly contradict the conventional 
polarization data of Nb - Co. 
On the other hand, Nb is claimed  \cite{Kerchner1981} to have a Fermi velocity 
of only $v_F = 273\,$km/s, based on  critical field measurements, with a 
heat-capacity derived effective mass enhancement of about 2. 
That would mean a perfect match between Nb and Co with $Z_0 \approx 0$.

Quite different estimates for $Z_0$ come from proximity-effect studies on 
Nb - normal metal bi-layers \cite{Cirillo2004,Tesauro2005,Attanasio2006}
with interface transparencies $1/(1+Z^2)$  consistently smaller than 50\%.
Since those bi-layers should have no (oxide) tunneling barrier, their Fermi 
surface mismatch must be large with $Z_0 \ge 1$ for non-magnetic normal 
metals Cu, Ag, Al, and Pd as well as for the ferromagnets Fe and Ni. 
The same is to be expected for Nb - Co interfaces \cite{Liang2002}.
This is difficult to reconcile with the standard interpretation of 
Andreev-reflection spectroscopy of the ferromagnets - here lifetime effects 
would fit much better.

If we assume that our Nb-Cu contacts are non-magnetic, then they deliver
the normal reflection $Z_0 \approx 0.8$ due to Fermi surface mismatch
in good agreement with the above mentioned proximity-effect data
where a tunneling barrier can be excluded.
The scattering $\Delta Z \approx \pm 0.2$ around the average could result, for
example, from small residual oxide barriers or the different crystallographic
orientations of the polycrystalline electrodes when the contacts are formed.
There is little reason to assume that Nb-Co contacts should have a much 
smaller Fermi surface mismatch even down to $Z_0 \approx 0$. 
The $P(Z)$ data points of Nb-Co at small $Z$ are therefore invalid. 
Shifting them to higher $Z$ values requires the inclusion of lifetime effects,
a quite natural consequence since we would expect the interface with 
ferromagnetic Co not to be less pair breaking than the one 
with non-magnetic Cu.
However, without precise knowledge of $Z$ it is difficult to extract any 
reliable value of the polarization. 
Our data even show that the Nb-Co contacts could be non-magnetic like the
Nb-Cu contacts. 
A small polarization at contacts with a large $Z$ would be consistent 
with predictions of the conventional theory \cite{Kant2002}. 

On the other hand, we can not exclude that Nb - Cu contacts are magnetic.
The Andreev-reflection spectra are consistent with a small local polarization
of Cu as has been suggested in Ref. \cite{Gloos2009}.

We have obtained similar Andreev-reflection data for the ferromagnets 
Fe and Ni as well as the non-magnets Ag and Pt in contact with Nb,
indicating a rather general problem of Andreev-reflection spectroscopy.

%#########################################################################
%#########################################################################
\section{Conclusions}

The available information suggests that the true (spin current) polarization 
of the ferromagnets is probably not that derived from Andreev-reflection 
spectra when lifetime effects are arbitrarily excluded
and the intrinsic normal reflection due to Fermi surface mismatch ignored. 

\begin{acknowledgments}
  We thank the Jenny and Antti Wihuri Foundation for financial support.
  \end{acknowledgments}

%\begin{acknowledgments}
% Put your acknowledgments here.
%\end{acknowledgments}

%\bibliography{references}

\end{document}